\documentstyle[prd,aps,twocolumn]{revtex}

\tighten

\begin{document}
\draft
 
\pagestyle{empty}

\preprint{
\noindent
\today \\
\hfill
\begin{minipage}[t]{3in}
\begin{flushright}
LBL--xxxxx \\
UCB--PTH--96/xx \\
hep-ph/96xxxxx \\
\end{flushright}
\end{minipage}
}

\title{Search of $^1P_1$ charmonium in $B$ decay}

\author{
Mahiko Suzuki
}
\address{
Department of Physics and Lawrence Berkeley National Laboratory\\
University of California, Berkeley, California 94720
}


\date{\today}
\maketitle

\begin{abstract}

There is no doubt that the $^1P_1$ chamonium $h_c$ exists in 
the mass range between $J/\psi$ and $\psi'$. While experiment 
produced a candidate in the past, it still requires a confirmation.  
Given the recent progress in the exclusive $B$ decay into 
charmonia, we now have an opportunity to detect $h_c$ by 
measuring the final state $\gamma\eta_c$ of the cascade decay 
$B\to h_c K/K^*\to\gamma\eta_c K/K^*$. Confirmation of $h_c$ 
may turn out to be much easier in the $B$ decay than at charm 
factories although one may have to work a little harder to attain 
a high precision in the mass determination. 

\end{abstract}
\pagestyle{plain}
\narrowtext

\setcounter{footnote}{0}

\section{Introduction}

   A few measurements suggested the $^1P_1$ charmonium 
around mass 3526 MeV\cite{Baglin,760,705}. In particular, the E760 
Collaboration\cite{760} studied the resonant production of $h_c$
in $p\overline{p}$ annihilation\footnote{
Although it is an odd naming, I call the $^1P_1$ charmonium as $h_c$ 
following the Particle Data tabulation\cite{PDG}.}
and quoted the $h_c$ mass at 3526.2 MeV. This value is almost exactly 
equal to the center of gravity (3525.17 MeV) of the $^3P_J$ charmonia 
$\chi_{cJ}$ ($J=0,1,2$). However, the result has yet to be 
confirmed by the E835 Collaboration\cite{835}. No evidence
has so far been seen for $h_c$ in $e^+e^-$ annihilation. From
the theoretical viewpoint, there is no reason to expect that the $h_c$ 
mass should be so close to the center of gravity of the $^3P_J$ masses,   
since such a relation based on the ${\bf L}\cdot{\bf S}$ coupling and
the tensor force of one-gluon exchange would break down when general 
spin-dependent interactions are included. Experimentally, 
the $\chi_{cJ}$ mass splitting gives
\begin{equation}
 R\equiv \frac{m_{\chi_{c2}}-m_{\chi_{c1}}}{m_{\chi_{1c}}-m_{\chi_{c0}}}
  \simeq 0.476.  
\end{equation}
The right-hand side would be equal to 2 for the ${\bf L}\cdot{\bf S}$ 
coupling alone, 0.8 with all spin-dependent forces of one-gluon exchange, 
and $0.8 \leq R \leq 1.4$ after including the more general spin-spin
interaction arising from the confining potential\cite{Schnitzer}. 
Since our knowledge of the spin-dependent charmonium potential 
is incomplete, there is no accurate theoretical prediction of 
$m_{h_c}$ relative to $m_{\chi_{cJ}}$ even within the potential 
model. Furthermore, the E1 transition 
matrix elements for $\chi_{cJ}\to\gamma J/\psi$ deviate 
largely from the nonrelativistic values. When relativistic 
corrections are large for the motion of $c$ and $\overline{c}$, 
we should be cautious about accuracy of the potential model approach.

Review of Particle Physics\cite{PDG} has not yet listed $h_c$ 
among the confirmed particles. Undoubtedly, much effort will be 
devoted to pursuit of $h_c$ at upcoming charm factories overcoming 
the odds against it. Meanwhile, the recent progress in $B$ physics 
suggests a new opportunity to search for $h_c$. The purpose of 
this short note is to point out that we may be able to observe 
$h_c$ more easily at the B-factories than at future charm factories 
and in hadron reactions. 

Recently the Belle Collaboration discovered that the 
factorization-forbidden decay $B\to\chi_0 K$ occurs as vigorously 
as the factorization-allowed decays to other charmonia\cite{Belle}. 
On the basis of this finding, we expect that another 
factorization-forbidden decay $B\to h_c K$ may also occur just 
as abundantly as $B\to \chi_{c0} K$. Since $h_c\to\gamma\eta_c$ 
is one of the two main decay modes of $h_c$, the decay $B\to h_c K$ 
cascades down to the final state $\gamma\eta_c K$ about half 
of time. The only background for this process 
at the $B$-factories will be the process $B\to \psi'K\to 
\gamma\eta_c K$. Since the branching fraction for 
$\psi'\to\gamma\eta_c$ is minuscure, however, this background is 
two orders of magnitude smaller than the signal. If one can 
reconstruct $\eta_c$ from $K\overline{K}\pi$ or by $\eta\pi\pi$ 
with 50\% efficiency, for instance, 10 million $B$'s translate 
to roughly 100 events of the signal. Therefore we have a very good 
chance to observe $h_c$ through $B\to\gamma\eta_c K$.    
  
\section{$B\to$ charmonium + $K$}

  The Belle Collaboration reported for the decay
$B\to\chi_{c0}K$\cite{Belle}
\begin{equation}
   {\rm B}(B^+\to \chi_{c0}K^+) = 
       (8.0_{-2.4}^{+2.7} \pm 1.0\pm 1.1) \times 10^{-4}.\label{Belle}
\end{equation}
This number should be compared with the recent measurement by the
BaBar Collaboration on the $B$ decay into other charmonia\cite{BaBar}:
\begin{eqnarray}
 {\rm B}(B^+\to J/\psi K^+) &=& (10.1\pm 0.3 \pm 0.5) \times 10^{-4}, 
                         \nonumber \\ 
 {\rm B}(B^+ \to\chi_{c1} K^+) &=& (7.5\pm 0.8\pm 0.8) \times 10^{-4},
                     \nonumber \\
 {\rm B}(B^+\to \psi' K^+) &=& (6.4\pm 0.5\pm 0.8)\times 10^{-4}. 
\end{eqnarray}
Added to these is an earlier measurement on the branching fraction 
for $B\to\eta_c K$ by CLEO\cite{CLEO}:
\begin{equation}
      B(B^+ \to \eta_c K^+) =
       (6.9_{-2.1}^{+2.6}\pm 0.8\pm 2.0)\times 10^{-4}.
\end{equation}
Most recently, however, BaBar gave a preliminary result for this
decay as\cite{BaBar2}
\begin{equation}
       B(B^+ \to \eta_c K^+) =
       (15.0 \pm 1.9 \pm 1.5 \pm  4.6)\times 10^{-4}. 
\end{equation}
We should notice here that the decay $B\to\chi_{c0}K$ is forbidden 
by the factorization while $B\to J/\psi(\psi')K$, $B\to\eta_c K$,
and $B\to\chi_{c1}K$ are all allowed. Nonetheless the branching 
fraction to $\chi_{c0}K^+$ is just as large as those into 
$J/\psi(\psi') K^+$, $\eta_c K^+$, and $\chi_{c1}K^+$. Since
no effective decay operators allows $B\to\chi_{c0}K$ in the
factorization limit, its decay amplitude must arise from 
the loop corrections of the energy scale below $m_b$ to the 
tree-decay operators ${\cal O}_{1,2}$. The relevant $\overline{c}c$ 
operator for production of $\chi_{c0}$ is generated when
the bilocal operator $\overline{c}(x)c(y)$ due to the loop 
correction is expanded in the series of local operators;
$\overline{c}(x)c(y)\rightarrow \overline{c}(x)c(x) + 
\overline{c}(x)(y-x)_{\mu}\partial^{\mu} c(x)+\cdots$.
If the relevant part of the loop energy is between $m_b$ and $m_c$,
then $|y-x|\simeq 1/m_b\sim 1/m_c$ so that one may keep only the
leading term of the expansion. In this case the QCD coupling 
$\alpha_s/\pi$ would suppress the $B\to\chi_{c0}K^+$ decay 
branching by $(\alpha_s/\pi)^2$ though the suppression relative 
to the factorization-allowed processes may be
somewhat moderated by the color structure. The experimental 
fact that ${\rm B}(B^+\to\chi_{c0}K^+)$ is comparable with 
${\rm B}(B^+\to\chi_{c1}K^+)$ indicates that the 
factorization, even improved with perturbative QCD corrections, 
is in serious doubt for the $B$ decay into charmonia.
In terms of the local expansion of $\overline{c}(x)c(y)$,
the magnitude of the relevant $|y-x|$ is as large as 
$1/\Lambda_{QCD}$ or, in the case of charmonia, could be 
the charmonium radius $1/\alpha_sm_c$. If so, we must include 
not only all terms of the local expansion but also all orders 
of $\alpha_s$ in computation of decay amplitudes. Then a 
quantitative calculation based on perturbative QCD is intractable.            

The decay $B\to h_c K$ is also forbidden by the factorization and has
the same chiral structure ($\overline{c}_Lc_R\pm\overline{c}_Rc_L$)
for charmonium as $B\to\chi_{c0}K$. The local operator of the lowest 
dimension leading to the decay $B\to h_c K$ is $\overline{c}
\gamma_5\!\stackrel{\leftrightarrow}{\partial^{\mu}}\!c$\cite{Suzuki}.
When the factorization and perturbative QCD fail as proven by
the $B\to\chi_{c0}K$ decay rate, it is very likely that the decay 
$B\to h_c K$ occurs as abundantly as $B\to \chi_{c0}K$ and 
the factorization-allowed $B$ decays into charmonia.

A comment is in order for another factorization-forbidden decay,
$B\to\chi_{c2}K$.  The decay $B\to\chi_{c2}K$ occurs with 
$i\overline{c}\gamma_{\mu}\!\stackrel{\leftrightarrow}{\partial^{\nu}}\!c$. 
The Belle Collaboration did not see a signal of $B\to\chi_{c2}K$ 
with a statistical significance\cite{Belle}. However, since they 
searched $\chi_{c0,c2}$ by its $\chi_{c0,c2}\to\pi^+\pi^-$ and $K^+K^-$
decay modes, their failure to see a clear signal for $B\to\chi_{c2}K$ 
may be due to the smaller branching fractions for $\chi_{c2}\to 
\pi^+\pi^-$ and $K^+K^-$ as compared with $\chi_{c0}\to\pi^+\pi^-$
and $K^+K^-$. On the other hand the CLEO Collaboration identified 
$\chi_{c2}$ by $\chi_{c2}\to J/\psi\gamma$ and concluded that 
$B(B\to\chi_{c2}X)$ is significantly less than 
$B(B\to\chi_{c1}X)$. But they focused on the inclusive decays 
and the uncertainties were large for the exclusive decays: 
$0.04 < B(B\to\chi_{c2}K/K^*)/B(B\to\chi_{c1}K/K^*) < 0.58$ with
the 95\% confidence level. (See the Sample B of Ref.\cite{CLEO2}.)
Very recently, however, the Belle Collaboration reported the 
branching fraction for inclusive $\chi_{c2}$ production\cite{Belle2},
\begin{equation}
  B(B\to\chi_{c2}X) = (15.3_{-2.8}^{+2.3}\pm 2.6)\times 10^{-4},
\end{equation}
where the $\chi_{c2}$'s fed by $\psi'\to\gamma\chi_{c2}$ have been 
subtracted out. This number is twice as large as the corresponding one
of CLEO\cite{CLEO2}. In view of this latest Belle measurement, 
it is possible that $B(B\to\chi_{c2}K)$ will eventually turn out to be
comparable with $B(B\to\chi_{c0}K)$.      
   
With these observations in theory and experiment, we proceed for
the moment with the assumption, 
\begin{equation}
     B(B\to h_c K) \approx B(B\to\chi_{c0}K) \label{equal}
\end{equation} 
to explore the opportunity to detect $h_c$. Once we assume 
Eq.(\ref{equal}), we are assuming the same relation with 
$K$ replaced with $K^*$ or a higher strange meson. 
We emphasize that Eq.(\ref{equal}) is an assumption at present.
However, the measurement we are duscussing will test its validity,
as we discuss below, and determine or set an upper bound on
$B(B\to h_c K)$ with a good accuracy.

\section{Decay of $^1P_1$}

 Numerous calculations were performed for the properties of
charmonia in potential models\cite{Schnitzer,Eichten}. 
The decay property of $\chi_{c1}$ and $h_c$ was specifically
studied by Bodwin {\em et al}\cite{Bodwin}. Production of 
$h_c$ through $\psi'\to h_c\pi^0$  in $e^+e^-$ annihilation was 
also studied\cite{Yan,Andri,Voloshin}. However, all that we need 
for our purpose here can be obtained directly from the experimental 
numbers for other charmonia if we make the approximation to use 
a common orbital wave function for the spin singlet and triplet of
the same principal quantum number. 
This approximation is justified for the $c$ and $\overline{c}$
in nonrelativistic motion, and the results are independent of specific 
bound-state wave functions. Although the nonrelativistic treatment 
of charmonia is often limited in precision, we do not need much
more than that for our discussion below.
        
   The main decay modes of $h_c$ are $h_c\to ggg$ and 
$h_c\to\gamma\eta_c$. The former is given by perturbative QCD 
to the leading logarithm of the $h_c$ size\cite{Barbieri}. 
By equating the $h_c$ bound-state wave function at the origin 
to that of $\chi_{c1}$, we obtain with the experimental value
$\Gamma(\chi_{c1}\to ggg)= \Gamma(\chi_{c1}\to hadrons)=640\pm 100$ keV,
\begin{eqnarray}
 \Gamma(h_c\to ggg) &=& \frac{5}{6}\times
          \Gamma(\chi_{c1}\to ggg), \nonumber \\
                    &=& 530\pm 80\; {\rm keV}. \label{ggg}
\end{eqnarray}  
This numerical value does not depend on the magnitude of the fuzzy 
cutoff variable in the leading logarithmic term nor 
on specific binding potentials.

 The radiative decay $h_c\to\gamma\eta_c$ is an allowed E1 transition 
similar to $\chi_{cJ}\to \gamma J/\psi$. We can eliminate the E1
transition matrix element $\langle f|{\bf r}|i\rangle$ between 
the $1P$ and the $1S$ state by relating $h_c\to\gamma\eta_c$ to 
$\chi_{c1}\to\gamma J/\psi$:
\begin{eqnarray}
  \Gamma(h_c\to\gamma\eta_c) &=& 
          \biggl(\frac{|{\bf p}|}{|{\bf p}'|}\biggr)^3
          \Gamma(\chi_{c1}\to \gamma J/\psi), \nonumber \\
           & = & 520 \pm 90 \;{\rm keV}. \label{gamma}
\end{eqnarray}
The central value of Eq.(\ref{gamma}) is about 15\% higher than 
the value computed by Bodwin {\em et al}\cite{Bodwin}, while the 
value $530 \pm 80$ keV of Eq.(\ref{ggg}) coincides with theirs.
The rates for other modes such as $h_c\to J/\psi\pi^0$ and 
$\gamma\chi_{c0}$ are of $O(1)$ keV. Therefore we obtain from
the $ggg$ and $\gamma\eta_c$ decay modes the $h_c\to\gamma\eta_c$ 
branching fraction;
\begin{equation}
  B(h_c\to \gamma\eta_c) =  0.50 \pm 0.11. \label{Br}
\end{equation} 
In this estimate the uncertainty is entirely due to those of the
measured values for
$\Gamma_{tot}(\chi_{c1})$ and $B(\chi_{c1}\to \gamma J/\psi)$.
The value of Eq.(\ref{Br}) is a firm number up to relativistic 
corrections and higher-order QCD corrections though the former 
corrections may turn out to be larger than we imagine. 

   Combining $B(h_c\to\gamma\eta_c)$ of Eq.(\ref{Br}) with 
Eqs.(\ref{Belle}) and (\ref{equal}), we obtain the cascade 
branching fraction for $B\to h_c K\to \gamma\eta_c K$;
\begin{eqnarray}
      B(B^+\to h_c K^+ &\to & \gamma\eta_c K^+) \nonumber \\
      &=& (4.0_{-1.5}^{+1.6}\pm 0.5\pm 0.6)\times 10^{-4}. \label{hc}
\end{eqnarray} 
It goes without saying that the number on the right-hand side is
subject to the uncertainty of the assumed equality in Eq.(\ref{equal}).
If $\eta_c$ is searched by $K\overline{K}\pi$ or $\eta\pi\pi$
(the branching fraction $\simeq 5\%$ each), the cascade branching 
fraction is  
\begin{eqnarray}
 B(B^+\to h_c K^+\to\gamma\eta_cK^+&\to&\gamma(K\overline{K}\pi)K^+)
                           \nonumber \\
             &\simeq& 2 \times 10^{-5}. \label{events}   
\end{eqnarray}
When 10 millions of $B$ mesons are accumulated, there will be
about 100 events of the $\gamma\eta_c K^+$ signal just from
$K\overline{K}\pi$ or from $\eta\pi\pi$ alone in the case that 
the reconstruction efficiency of $h_c$ is 50\% for these decay modes. 
One can increase statistics by including $B^{\pm}\to h_c K^{*\pm}$  
and by combining $B^0/\overline{B}^0$ with $B^{\pm}$. There will
be a sufficient number of the cascade $B\to h_c X \to \gamma\eta_c X$ 
events to search for $h_c$.

Let us compare Eq.(\ref{events}) with the corresponding number in
the $h_c$ search through $\psi'\to\gamma h_c$ at charm factories.  
According to the calculation by Yan {\em et al}\cite{Yan} and more 
recently by Kuang\cite{Kuang} who included $S$-$D$ mixing of $\psi'$, 
the branching fraction for $\psi'\to h_c\pi^0$ is at the level of 
$1\times 10^{-3}$ at most, for $m_{h_c}= 3526.2$ MeV. Taking 
account of the low reconstruction efficiency of the soft 
$\pi^0\to\gamma\gamma$, Kuang estimates that detection of $h_c$ 
through $\psi'\to h_c\pi^0$ requires 30 million $\psi'$'s at charm 
factories. While $h_c$ can be produced only through 
$\psi'\to h_c\pi^0$ at charm factories, the $h_c$ production 
occurs in the $B$ decay in conjunction with $K^*$ or a higher 
strange meson as well as with $K$. Furthermore, the production 
in conjunction with $K^*$ tends to be stronger than that
with $K$ in the $B\to charmonium$ decay. By and large, the search of 
$h_c$ will be quite competitive with the search at charm 
factories, if not superior to it. 

\section{Possible background events}

 The only decay mode that feeds $\gamma\eta_c K$ with the
$\gamma\eta_c$ invariant mass close to $m_{h_c}$ is the cascade 
decay $B\to\psi'K\to\gamma\eta_c K$. Since $\psi'\to\gamma\eta_c$ 
is a hindered M1 transition with the branching fraction 
$(2.8\pm 0.6)\times 10^{-3}$, this cascade branching fraction is tiny;
\begin{equation}
  B(B\to\psi' K^+\to\gamma\eta_c K^+)
  = (1.8 \pm 0.4 \pm 0.2)\times 10^{-6}.
\end{equation} 
It is more than two orders of magnitude smaller than the signal
of Eq.(\ref{hc}).  We can therefore choose a wide bin for
$(p_{\gamma}+p_{\eta_c})^2$ in reconstruction of $h_c$ without concern 
about the $\psi'$ contamination in $\gamma\eta_c$. This is fortunate 
from the viewpoint of raising the precision in mass determination. 
Since there is no competing decay process, we may fix the invariant 
mass of $K\overline{K}\pi$ or $\eta\pi\pi$ to $m_{\eta_c}$ once
we find a cluster of candidate events. Although we certainly do 
not expect to determine the $h_c$ mass to the accuracy 
anywhere close to its width ($\Gamma_{h_c}\simeq 1$ MeV), 
it will be easy to notice if the $h_c$ mass is located
substantially off the center of gravity of $\chi_{cJ}$.

It will be challenging to identify $h_c$ directly by its hadronic
decay modes. Since $h_c$ is G-parity odd, the simplest decay mode is  
$h_c\to\pi\pi\pi$, then $K\overline{K}\pi$. The branching fractions 
to $\pi\pi\pi$ and $K\overline{K}\pi$ are no larger than at the level 
of 1\% if we make a guess by rescaling the corresponding decays for 
$\psi'$. Then the cascade branching fraction is most likely of 
the order,
\begin{equation}
  B(B\to h_c K\to \pi^+\pi^-\pi^0 K) = O(1) \times 10^{-5}.
\end{equation}
After multiplying it with the reconstruction efficiency of 
$\pi^0\to\gamma\gamma$, it does not appear competitive with 
$B\to h_c K\to \gamma\eta_c K$. Although one can distinguish $h_c$ 
from $\chi_{c1}$ by G-parity of the decay products, one can separate 
$h_c$ from $\psi'$ only by the mass resolution when one searches
$h_c$ by it hadron decays. There are clear advantages for studying 
the cascade decay $B\to h_c K \to \gamma\eta_c K$.

\section{Summary}
 
 Nobody disputes the presence of $h_c$. Our real interest is in 
the values of its parameters. For this purpose the cascade decay 
process $B\to h_c K/K^*\to \gamma\eta_c K/K^*$ deserves a careful 
study at the B-factories.  The search of $h_c$ through the $B$ 
decay is very competitive with the search at charm factories and
presumably superior to it. It will either confirm the controversial 
$^1P_1$ charmonium at the center of mass gravity of $\chi_{cJ}$ or
discover it off the value suggested by $p\overline{p}$ annihilation. 
We should keep in mind that theory does not require 
that the $h_c$ mass should be so close to the center 
of gravity of $\chi_{cJ}$. 

We shall obtain the product of the branching fractions, 
$B(B\to h_c K)\times B(h_c\to\gamma\eta_c K)$ from the proposed 
$B$ decay measurement. Since the value of $B(h_c\to\gamma\eta_c)$ 
given in Eq.(\ref{Br}) is a fairly firm number, measurement or 
nil measurement of the process $B\to h_c K \to \gamma\eta_c K$ 
will provide us with a meaningful number or a tight upper bound 
for $B(B\to h_c K)$.
If we end up with a nil result for $B\to h_c K \to \gamma\eta_c K$,
it would mean that $B(B\to h_c K)$ is for some reason much 
smaller than $B(B\to\chi_{c0}K)$. Whatever the experimental 
outcome will be, such information will provide us with an opportunity 
to examine all of $B\to charmonium$ decays together and will 
advance our understanding of how or if the factorization plays 
a role in the $B$ decay into charmonia.

\section{Acknowledgment}
I am benefited from conversations with R.N. Cahn on the mass 
resolution in the BaBar data analysis.
This work was supported in part by the Director, Office of Science, 
Office of High Energy and Nuclear Physics, Division of High Energy 
Physics, of the U.S.  Department of Energy under contract 
DE-AC03-76SF00098 and in part by the National Science Foundation 
under grant PHY-0098840.

\end{document}